\documentclass[pra,amssymb,twocolumn]{revtex4}
\usepackage{graphicx,amsbsy,amsmath,bbm}
\usepackage[usenames]{color}

\newcommand{\trace}{{\mathrm Tr }}

\newcommand{\twomat}[4]{\left(\begin{array}{cc} #1 & #2 \\ #3 & #4\end{array}\right)}

\newcommand{\identity}{\openone}
\newcommand{\be}{\begin{equation}}
\newcommand{\ee}{\end{equation}}
\newcommand{\bea}{\begin{eqnarray}}
\newcommand{\eea}{\end{eqnarray}}
\newcommand{\beas}{\begin{eqnarray*}}
\newcommand{\eeas}{\end{eqnarray*}}
\newcommand{\refeq}[1]{Eq.~(\ref{#1})}
\newcommand{\eg}{{\it{e.g.}}}
\newcommand{\ie}{{\it{i.e.}}}

\begin{document}
\title{Monogamy and ground-state entanglement in highly connected systems}
\author{Alessandro Ferraro, Artur Garc\'\i a-Saez and Antonio Ac\'\i n} 
\affiliation{ICFO-Institut de Ciencies Fotoniques, Mediterranean
  Technology Park, 08860 Castelldefels (Barcelona), Spain }
\begin{abstract}
  We consider the ground-state entanglement in highly connected
  many-body systems, consisting of harmonic oscillators and
  spin-$1/2$ systems. Varying their degree of connectivity, we
  investigate the interplay between the enhancement of entanglement,
  due to connections, and its frustration, due to monogamy
  constraints. Remarkably, we see that in many situations the degree
  of entanglement in a highly connected system is essentially of the
  same order as in a low connected one. 
  We also identify instances in which
  the entanglement decreases as the degree of connectivity increases.
\end{abstract}
\date{\today}
\maketitle
\section{Introduction}
Entanglement theory has experienced an impressive development in
the last decade, mainly due to the key role quantum correlations
play in quantum information science. 
The novel concepts and mathematical methods developed in this new
research area are beginning to reveal their usefulness also in
different contexts. A striking example of this tendency is given
by the physics of quantum many-body systems. As an instance, the
analysis of the role played by entanglement in quantum phase
transitions allowed for a deeper understanding of this purely
quantum phenomenon \cite{OAFF02,ON02,VLRK03}. In this scenario,
entanglement theory is also giving a fundamental contribution in
the development of new methods capable of simulating efficiently
strongly interacting systems \cite{V04,VPC04,APD+06}.

Clearly, the correlations between different parts of a many-body
system originates by their mutual interaction. In this sense
it is natural to expect that the ground state of a strongly
interacting and connected quantum system will exhibit a high degree of
entanglement. However, this intuition has to be taken cautiously,
since the shareability properties of quantum correlations are
especially non trivial and without classical analogue. One of the main
differences between classical and quantum correlations is the so
called monogamy of the latter \cite{CKW00}. In the classical scenario,
the fact that two systems share some correlations does not prevent
them from being correlated with a third party. On the contrary, two
maximally entangled quantum systems can share no correlation at all
with a third one. More generally, quantum correlations are not
infinitely sharable, and the more the entanglement the less the number
of systems with which it can be shared.

Consider two similar Hamiltonians consisting of the same
interacting terms between pair of particles, the only difference
being the degree of connectivity. One of them, for instance, has
only nearest-neighbor interactions, while the second has also
next-to-nearest-neighbor interactions. Let us focus on the
ground-state entanglement between two halves of the system.
Naively, the more connected hamiltonian is expected to have a
larger entanglement, since there are more bonds connecting the two
halves. However, in the more connected system, each particle has
to share the quantum correlations with a larger number of
particles, so the connecting bonds may give a smaller amount of
entanglement. Therefore, it is unclear which geometry leads to a
larger ground-state entanglement.

In this work we analyze the interplay between the enhancement of the
ground-state entanglement due to connections and its suppression due
to monogamy constraints. We consider spin-$1/2$ and infinite
dimensional (harmonic oscillators) systems of various geometries
with two-body interactions and focus on the bipartite entanglement
between two halves of the system.  Remarkably, we see that in many
situations the degree of entanglement in a highly connected system
is essentially of the same order as in the case of a low connected
one. Actually, we can even individuate systems for which the
entanglement decreases as the degree of connectivity increases.

Before proceeding, let us mention that there exist some works
studying how the monogamy of entanglement affects the ground state
properties of Hamiltonians with nearest-neighbor interactions, see
for example \cite{OW01,WVC04}. 
The ground-state entanglement
of a highly symmetric and connected system, the so-called
Lipkin-Meshkov-Glick model, was also computed in \cite{LOR+05}.

The paper is organized as follows. In the next Section we will
introduce the systems we are going to analyze, in particular recalling
known results about entanglement calculations in the different cases
considered. Then, in Section \ref{geometries}, we will show in details
how the entanglement depends on the connectivity of the systems
themselves. In order to individuate a general behavior we will
consider a variety of different connectivity conditions. Specifically,
we will analyze both regular and random graph configurations, paying
particular attention to the case of bipartite graphs. The relationship
between our findings and related results in the context of
entanglement-area laws will be outlined in Sec.\ref{arealaw}. Even if
a way to quantify the action of monogamy in a multipartite setting is
still lacking, we will see in Sec.\ref{monogamy} that the analysis of
bipartite monogamy inequalities can shed some light onto our findings.
We will close the paper in Sec.\ref{esco} by discussing possible
implications of our findings, in particular in the context of
classical simulations of quantum systems.

\section{Spin and bosonic models}
As said, we consider two paradigmatic systems, namely interacting
spin-1/2 and bosonic particles. Concerning the former, we study a
system of $n$ spin-$1/2$ particles under the XX Hamiltonian, 
\be
\label{HXY}
H = \sum_{i,j} t_{ij}[\sigma_x^i\otimes\sigma_x^j +
\sigma_y^i\otimes\sigma_y^j],
\ee
where $\sigma_k^i$ ($k=x,y,z$) denote the Pauli matrices referred to
the $i$-th particle.  The coupling $t_{ij}$ will be set different from
zero when the $i-j$ couple directly interacts, \ie~in dependence on
the topology and connectivity of the system. The actual value of the
nonzero $t_{ij}$ will be chosen randomly, in order to have averaged
properties and avoid the dependence of our results on the details of
the interaction. Interactions of the type~(\ref{HXY}) may model
highly connected physical systems, such as quantum spin glasses
\cite{SpGl}. The entanglement between two parts of the system will
be measured by the entropy of entanglement $E$, namely the von Neumann
entropy $S(\rho)=-\trace[\rho\,\log_2\rho]$ of one of the reduced
subsystems \cite{BBPS96}.

Concerning the bosonic case, we consider systems consisting of $n$
harmonic oscillators with quadratic coupling. Such systems may model
discrete versions of Klein-Gordon fields, or vibrational modes in
crystal lattices, ion traps and nanomechanical oscillators.  We define
the vector $R$ of quadrature operators by $R_j=\hat{X}_j$ and
$R_{n+j}=\hat{P}_j$ ($1\le j\le n$), where $\hat{X}_j$ and $\hat{P}_j$
are the position and linear momentum operator respectively. For
simplicity, we consider only a coupling via the different position
operators, in which case the Hamiltonian is of the form 
\be \label{Hosc}
\hat{H} = R^T \twomat{V/2}{0}{0}{\identity_{n}/2} R, 
\ee 
where $\identity_{n}$ denotes the
$n\times n$ identity matrix. The potential matrix $V$ is
defined via the harmonic coupling between oscillator $i$ and $j$,
namely $\alpha(\hat{X}_i-\hat{X}_j)^2/2$.  For each geometry
considered in the following, we denote by $C$ the $n\times n$
adjacency matrix of the corresponding graph, with elements
$c_{ij}=c_{ji}=1$ if the $i$-th and $j$-th oscillator are coupled and
$c_{ij}=c_{ji}=0$ otherwise.  Then, the potential matrix $V$ is given
by $V_{ij}=-\alpha c_{ij}$ ($i\ne j$) and $V_{ii}=1+\alpha\sum_{j}
c_{ij}$. The ground state of the system is a Gaussian state
characterized by the covariance matrix $\gamma =
(\gamma_x\oplus\gamma_p)/2$, with $\gamma_x = V^{-1/2}$ and $\gamma_p
= V^{1/2}$ 
\cite{AEP+02}. We use as entanglement measure the
logarithmic negativity \cite{VW02}, $N_l$, between two generic group
$A$ and $B$. It can be shown that $N_l$ is given by  \cite{AEP+02}
\be
N_l = -\sum_{j=1}^{n}
\log_2\min[1,\Lambda_j(\gamma_x P \gamma_p P)],
\ee
where $\Lambda_j(M)$
is the $j$-th eigenvalue of matrix $M$. We denote by $P$ the $n\times
n$ diagonal matrix with $j$-th diagonal entry given by $1$ or
$-1$, depending on whether the oscillator on position $1\leq j\leq
n$ belongs to group $A$ or $B$, respectively.

\section{Entanglement in different topologies}\label{geometries}

\subsection{Neighbor coupling}\label{g:nc}
\begin{figure}[h!]
\includegraphics[width=8.8cm]{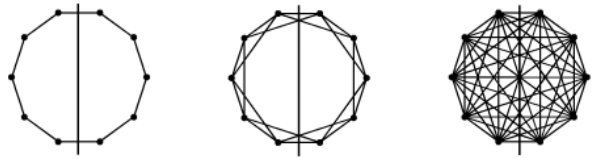}
\caption{\label{fig:graph}  Closed chain of $n=10$ interacting particles in 
  the nearest-neighbor ($n_c = 1$, left), next-to-nearest neighbor
  ($n_c = 2$, center) and fully connected configuration ($n_c = 5$,
  right). The entanglement is computed between two halves of the
  system.}
\end{figure}

\begin{figure}[h!]
\includegraphics[width=7cm]{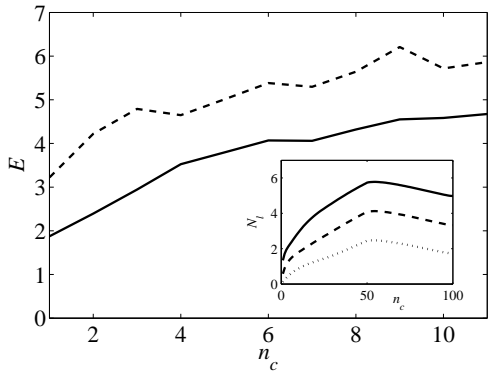}
\caption{\label{fig:n22}  For a closed chain of $n=22$ spin--$1/2$
  particles with XX interaction (\ref{HXY}), the entropy of
  entanglement is plotted versus the number of connected neighbors
  $n_c$, averaged over 100 realizations. The dashed line gives the
  largest entanglement obtained. Inset: for an open chain of $n=100$
  oscillators the logarithmic negativity $N_l$ is plotted versus
  $n_c$.  From top to bottom the coupling constant $\alpha$ is given
  by $\alpha=10,1,0.1$.}
\end{figure}

The first configuration that we consider is a one-dimensional (1D)
chain of $n$ particles, in which each of them can interact with $n_c$
of its neighbors on each side. Thus, $n_c$ is the parameter that
characterizes the degree of connectivity in this setting (see
Fig.~\ref{fig:graph}). We consider a distance-independent interaction,
in order to avoid any dependence on the particular scaling of the
interaction strength. In particular, given the ground state, we
calculate the entanglement between the two halves of the system
(groups $A$ and $B$) as a function of the number of interacting
neighbors $n_c$.
The typical behavior in the case of a XX system, is reported in
Fig.~\ref{fig:n22} for the case of $n=22$ spins. The exact calculation
of the ground state was performed using the SPINPACK package
\cite{spinpack}.  The solid line represents the averaged ground-state
entanglement, where the Hamiltonian parameters $t_{ij}$ between pair
of particles are randomly chosen in the interval $[0,1]$, while the
dashed line gives the largest entanglement obtained. We clearly see
that the entanglement grows only slightly and, in particular, the
fully connected chain has a degree of entanglement comparable to the
nearest-neighbor coupled chain. Note that, by contrast, the number of
bonds connecting the two halves of the chain increases as $n_c^2$. The
same behavior has been observed for different Hamiltonian operators,
consisting of other interaction terms, and smaller sizes.

We consider now the same configuration for the case of a chain of
harmonic oscillators. As said above, the interactions between the
particles simply correspond to oscillators of coupling constant
$\alpha$. The entanglement between the two halves of an open chain
consisting of $n=100$ oscillators is shown in the inset of
Fig.~\ref{fig:n22}, where the logarithmic negativity is plotted versus
the number of coupled neighbors, $n_c$. One clearly see that the
entanglement increases (almost linearly) as far as $n_c\lesssim n/2$,
whereas for higher connected systems the entanglement is frustrated.
The frustration mechanism is indeed stronger than in the spin case,
the entanglement decreasing at some point as the number of connections
increases. Notice the quite universal behavior of these curves: the
position of the maximum does not depend on the coupling constant
$\alpha$ and, as one can expect, the entanglement increases with
$\alpha$, for fixed $n_c$.

Both the examples reported here confirm that the monogamy of
entanglement plays a predominant role for highly connected
systems. As said, as the connectivity increases, each particle of,
{\em e.g.}, set $A$ becomes as well entangled with many other
particles of the same set. This in turn limits, for monogamy
reasons, the entanglement with the particles of set $B$.

\subsection{Random coupling} \label{g:rc}
\begin{figure}[ht]
\includegraphics[width=7cm]{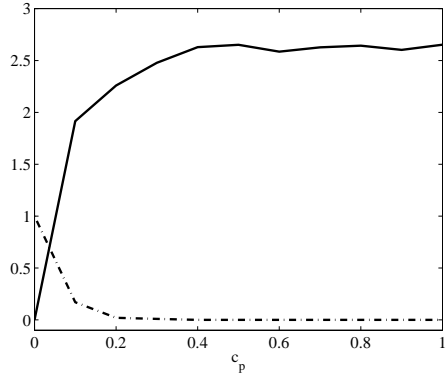}
\caption{\label{fig:ini12} Entropy of entanglement for a randomly coupled system with $n=10$ spins
  (solid), its degeneracy (dashed-dot) and its energy (dashed), as a
  function on $c_p$. We averaged over 100 executions.}
\end{figure}

We have seen above that the details of the entanglement frustration
mechanism depend on the system under consideration, nevertheless the
general behavior is the same for a variety of situations. As another
example, particularly different from the 1D chain above, consider now
a random configuration in which each couple of particles $i$ and $j$
is connected with probability $c_p$, which will be called connectivity
parameter.  Then, the connections are given by the edge of a graph
with random (symmetric) adjacency matrix. Concerning the case of the
XX system, the typical behavior is reported in Fig.~\ref{fig:ini12},
where the entropy of entanglement is plotted as a function of the
connectivity parameter. We also plot the degeneracy of the ground
state. Indeed, if the degeneracy was non-zero for a significant range
of values of $c_p$, the results should be carefully considered, since
there may be other ground states having quite different entanglement
properties. However, this does not affect the validity of our results,
since the degeneracy is equal to zero already for small values of the
connectivity parameter. Actually, this degeneracy disappears for
values of $c_p$ close to $n/n_{\rm t}$ (where $n_{\rm t}$ is the
number of connections of the complete graph), \ie~when we are able to
construct a connected graph with high probability. In general we see
again that the entanglement saturates, and the saturation value is
reached for low value of the connectivity, showing again the strong
effect of monogamy.

In the case of bosonic oscillators, one also retrieves the general
behavior already highlighted in the previous configuration.  Namely,
the entanglement increases up to a certain value of the connectivity
parameter ($c_p$ in the present configuration), after which the
monogamy constraint starts causing a decrease of the entanglement.

\subsection{Bipartite graphs} \label{g:bip}
\begin{figure}[h!]
\includegraphics[width=8cm]{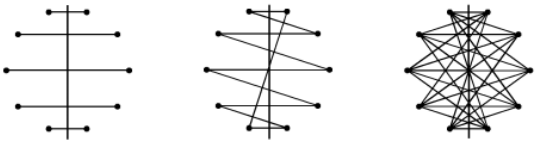}
\caption{Regular bipartite graph of $n=10$ interacting particles for
  increasing connectivity.}
\label{fig:bg}
\end{figure}

We exploit now the configuration in which perhaps the
effects of monogamy show up more impressively, namely the case in
which the system can be represented by a bipartite graph. The latter
is constituted by two sets, $A$ and $B$, for which particles belonging
to the same set do not directly interact (see Fig.~\ref{fig:bg}).  Then, a priori, one could
expect that in such a configuration the effects of monogamy should be
weaker than in the configurations discussed above. We report here the
results for both regular and random bipartite graphs in the case of
harmonic oscillators.

\begin{figure}[h!]
\includegraphics[width=8cm]{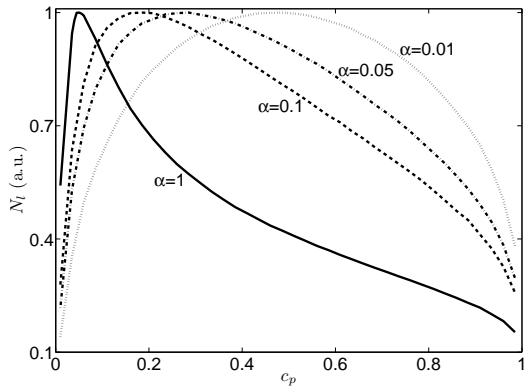}
\caption{For a random bipartite graph of harmonic oscillators the
  logarithmic negativity $N_l$ is plotted (normalized for each
  $\alpha$) versus the connectivity parameter $c_p$, for different
  coupling constants $\alpha$. The total number of particles is
  $n=100$. } \label{fig:tcg}
\end{figure}

Let us first consider the case of a random bipartite graph. As above,
we name connectivity parameter $c_p$ the probability that a generic
particle in $A$ interacts with a particle in $B$. For example in the
fully connected case ($c_p=1$) each particle in $A$ is coupled to all
the particles in $B$. For a fixed coupling constant we look for the
optimal $c_p^{\rm opt}$ such that the entanglement is maximized. As
shown in Fig.~\ref{fig:tcg}, we have that $c_p^{\rm opt}\neq 1$ in
general, depending non-trivially on $\alpha$.
Namely, for large values of $\alpha$ the maximum entanglement is
provided by an Hamiltonian with few connections for each oscillator.
Vice-versa, for low values of $\alpha$ the completely connected case
tends to maximize the entanglement. Note again that particles
belonging to the same set do not directly interact. However, these
particles become entangled through the common interaction with the
particles of the other set. This, at the same time, limits the amount
of entanglement between the two sets because of monogamy.

\begin{figure}[h!]
\includegraphics[width=8cm]{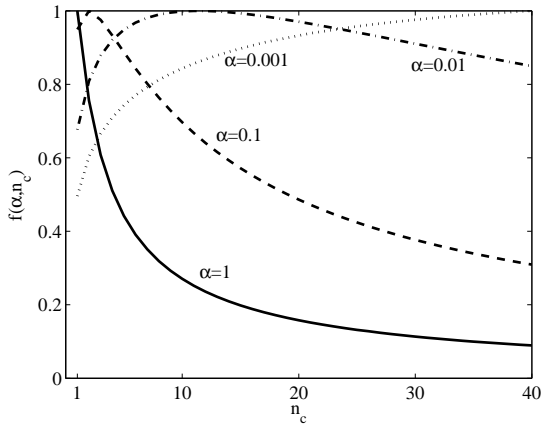}
\caption{For a regular bipartite graph of harmonic oscillators the function
  $f(\alpha,n_c)$ in \refeq{f_bg} is plotted versus the connectivity
  parameter $n_c$ (arbitrary units, normalized for each $\alpha$), for
  different coupling constants $\alpha$ (see text for details). }
\label{bg_regular}
\end{figure}

In order to obtain a deeper insight into the behavior described above,
let us now consider a regular graph. In this case, remarkably, it is
straightforward to obtain an explicit expression of the log-negativity
for an arbitrary large number of particles.  Let us focus on a
periodic system consisting of $n$ oscillators (labeled from $1$ to
$n$) and consider two sets composed such that the even oscillators
belong to $A$ and the odd ones to $B$. If each oscillator in set $A$
interacts only with $2n_c$ oscillators in $B$ (specifically, the
$2j$-th oscillator interacts with oscillators labeled $2k+1$ with
$k=j-n_c,\dots,j+n_c$), then the interaction matrix $V$ in
\refeq{Hosc} is given by a circulant matrix $V_{\rm bg}$, whose first
row is given by the following vector
\be {\bf v} = (1+2\alpha n_c,
\underbrace{-\alpha,0,\dots,-\alpha,0}_{n_c {\rm times}} ,0,\dots,0,
\underbrace{0,-\alpha,\dots,0,-\alpha}_{n_c {\rm times}}) 
\ee 
The eigenvalues of $V_{\rm bg}$ are given by
\be \lambda_k=1+2\alpha
n_c-2\alpha\sum_{j=1}^{n_c}\cos[(2j-1)k\frac{2\pi}{n}]\;, 
\ee 
hence, following Ref.~\cite{AEP+02}, the log-negativity between the
sets $A$ and $B$ can be straightforwardly obtained:
\be
N_l=\frac12\sum_{k=0}^{n/2-1}\left|
\log_2\frac{\lambda_k}{\lambda_{\frac{n}{2}-k}}\right|\;.
\ee
When $n$ is large, the sum above turns into a Riemann series,
which gives an explicit expression for the log-negativity: 
\be
N_l\approx\frac{n}{4\pi}f(\alpha,n_c)\;,
\label{nl_rbg}
\ee
where we have defined the function
\bea
f(\alpha,n_c)&=&\int_{0}^{\pi}\!\!\!\!\!dx\bigg|
\log_2\Big\{1+2\alpha n_c-2\alpha\sum_{1}^{n_c}\cos[(2j-1)x]\Big\}
\nonumber \\
&&\!\!\!\!\!\!\!\!\!\!\!\!\!\!\!\!\!\!\!\!\!
-\log_2\Big\{1+2\alpha n_c-2\alpha\sum_{1}^{n_c}\cos[(2j-1)(\pi-x)]\Big\}
\bigg|
\label{f_bg}
\eea 
in order to single out the dependence on $\alpha$ and the connectivity
parameter $n_c$. Noticeably, in \refeq{nl_rbg} the dependence on $n$
factories, allowing to analyze how the connectivity affects the
entanglement by simply looking at $f(\alpha,n_c)$ for different
coupling regimes (see Fig.~\ref{bg_regular}). It can be seen that the
results are very similar to the ones of a random bipartite graph,
indicating that the asymmetries of the latter do not play a
significant role. As in Fig.~\ref{fig:tcg}, for weak coupling the
entanglement increases with the connectivity of the system, meaning
that monogamy does not play a significant role. On the other hand, for
strong coupling the entanglement suddenly decreases with $n_c$. This
is a clear sign of the action of monogamy constraints, since for high
values of $\alpha$ a high degree of entanglement is created already
for small $n_c$.

\section{Connectivity and Area Law}\label{arealaw}
In this section we want to elucidate the relation between the findings
exposed above and the general issue concerning entanglement-area laws.
A number of theoretical investigations has shown that, for a variety
of physical systems, the entanglement between two complementary
regions scales as the area between them \cite{area}. For instance, it
is shown in Refs.~\cite{VLRK03,PED+05} that, for non-critical systems
with nearest-neighbor coupling, the entanglement between a
distinguished part of a system and the rest increases as the number of
connections.  Regarding more general interactions, in particular
long-range ones, the boundary area (given by the connections between
the two regions of the system) only gives an upper bound to the
entanglement \cite{CE06}.  In all these works the Hamiltonian of the
global system is kept fixed, whereas the size of the distinguished
region varies. It is then clear that the monogamy constraints do not
act significantly, since the connectivity of the systems remains
unchanged with the size. On the contrary, in most of our previous
analysis we kept fixed the size of the distinguished region, whereas
the connections between particles were modified by changing the degree
of connectivity of the Hamiltonians (see
Figs.~\ref{fig:n22},~\ref{fig:ini12} and \ref{fig:tcg}).
\begin{figure}[ht]
\includegraphics[width=7cm]{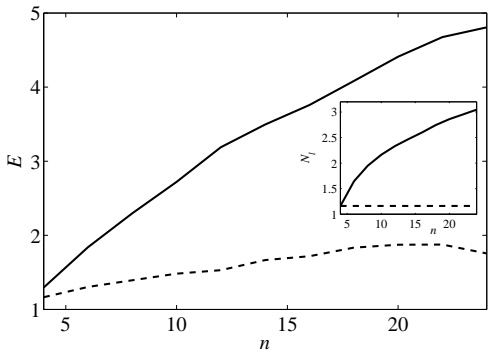}
\caption{\label{fig:quot} For an XX spin system
  the entropy of entanglement $E$ between the two halves of the system
  is plotted as a function of the system size $n$ (averaged over 100
  realizations).  Inset: corresponding graph for the negativity $N_l$
  in a closed harmonic chain with $\alpha=1$ for nearest-neighbor
  coupling (dashed line) and optimal coupling (solid line, see text).}
\end{figure}
\par
Following an approach suitable for a comparison with the works on area
laws, we exploited also the dependence of the entanglement on the size
of the system. The results are reported in Fig.~\ref{fig:quot}, where
the main panel refers to the case of an XX spin system and the inset
to an harmonic one.  In both cases we considered the non-random
configuration exploited in Sec.~\ref{g:nc}. In particular, we
focused on the results corresponding to {\em i}) nearest-neighbor
coupling and {\em ii}) the optimal configuration in which the number
of connections $n_c$ is chosen in order to give the maximal amount of
entanglement.  For a closed harmonic chain, we observe that the
entanglement remains constant in the nearest-neighbor case (as we
expect from the results of Ref.~\cite{AEP+02}), whereas it increases
only logarithmically in the optimally connected case. Actually, an
exact logarithmic increase can be derived in a fully connected
topology (\ie, $n_c=n$), for which the computation of the
log-negativity is straightforward following again Ref.~\cite{AEP+02}.
There, a lower bound to the log-negativity is given as a function of
the coupling parameters.  Fortunately, such bound turns out to be
tight in the case of both nearest-neighbor and fully connected systems
\cite{note}. In particular, for the latter one the log-negativity is
given by
\begin{equation}
N_l=\frac12 \log_2[1+2n\alpha]\;,
\end{equation}
proving the logarithmic increase with $n$. Remarkably, the behavior is
similar for a closed spin chain, as can be seen in
Fig.~\ref{fig:quot}. Recall that in this case the optimal number of
connections $n_c^{\rm opt}$ is always given by $n_c^{\rm opt}=n$, as
pointed out in Sec.~\ref{g:nc}.  Although our computations are not
very conclusive, they suggest a sub-linear increase in this case too.

As said, the scenario in Fig.~\ref{fig:quot} is suitable for a comparison with
the works concerning area laws, in particular the ones dealing with
half spaces (see, \eg, Ref.~\cite{CEP07}). We see that, as expected,
for nearest neighbor interactions an entanglement-area law is
satisfied (since we are not dealing with critical systems). On the contrary,
the slight entanglement increase for highly connected systems
strongly contrasts with the increase of the number of bonds linking
the two halves of a chain, which scales as $(n/2)^2$. As recalled, in
these cases the boundary area only gives an upper bound for the
entanglement. Thus, our results reveal that the entanglement can
actually scale sensibly slower than the area in highly connected
systems.

\section{Monogamy inequalities} \label{monogamy}
To quantify how the monogamy of entanglement acts in the analyzed
scenarios we can refer to monogamy inequalities.  In particular, for
the case of harmonic oscillators we considered the inequality
introduced in Ref.~\cite{HAI06}, where the Gaussian tangle $\tau$ is
used as entanglement measure. We recall that the latter is defined as
the square of the negativity, for pure states, and then extended to
mixed states via a convex roof. In particular, the monogamy constraint
for $n$ modes can be expressed as
\be
\label{g_monogamy}
\tau^{1:2\dots n}\ge\sum_{j=2}^{n}\tau_l^{1:j} 
\ee 
where $\tau^{1:2\dots n}$ denotes the tangle between the first mode
and the remaining ones, whereas $\tau_l^{1:j}$ is the two-mode tangle
between the first mode and the $j$-th one. The difference between the
lhs and the rhs of \refeq{g_monogamy} is called residual entanglement
and it indicates the presence of a complex structure of the quantum
correlations, not simply ascribable to two-mode entanglement.

\begin{figure}[h!]
\includegraphics[width=7cm]{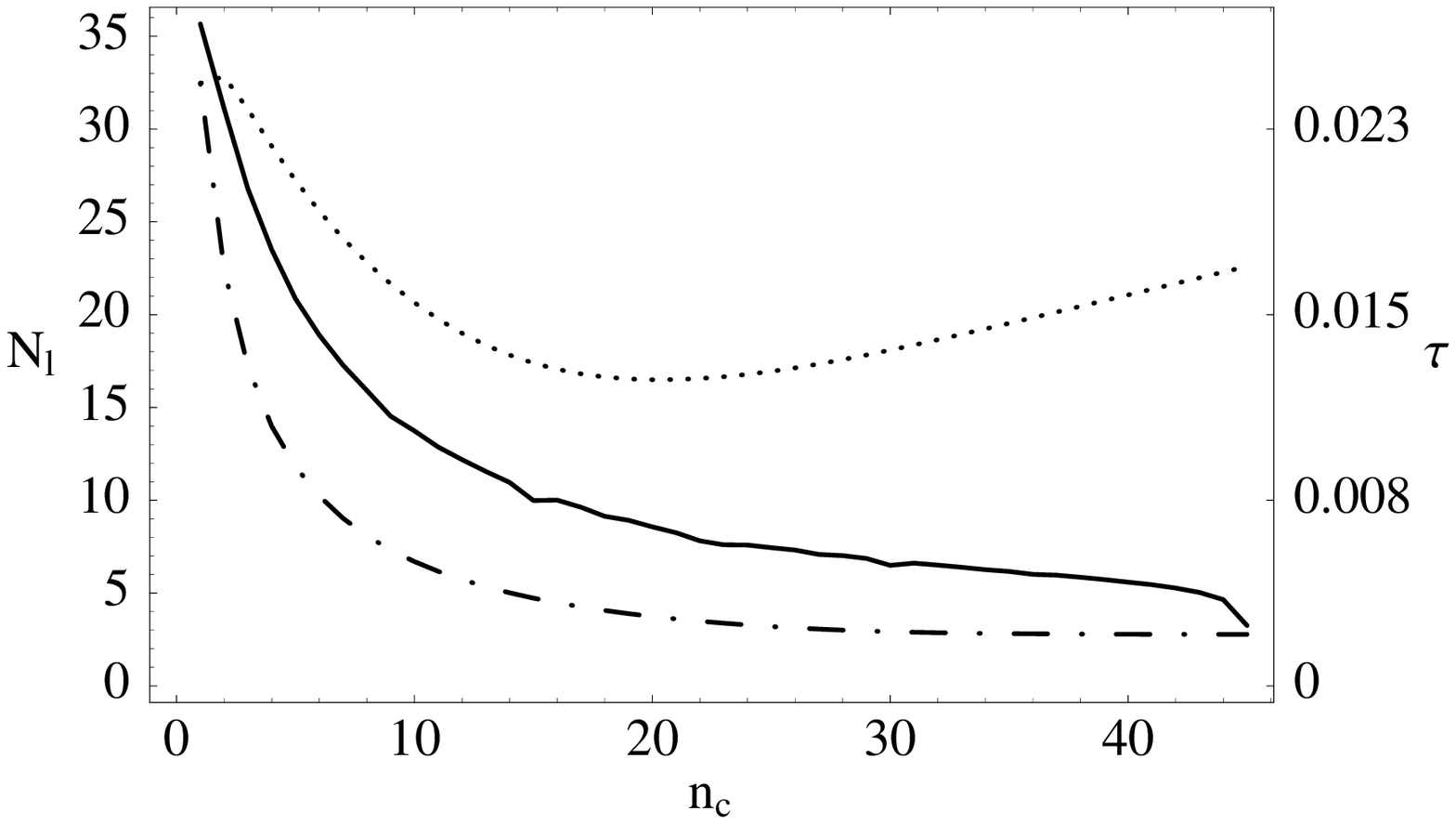}
\includegraphics[width=7cm]{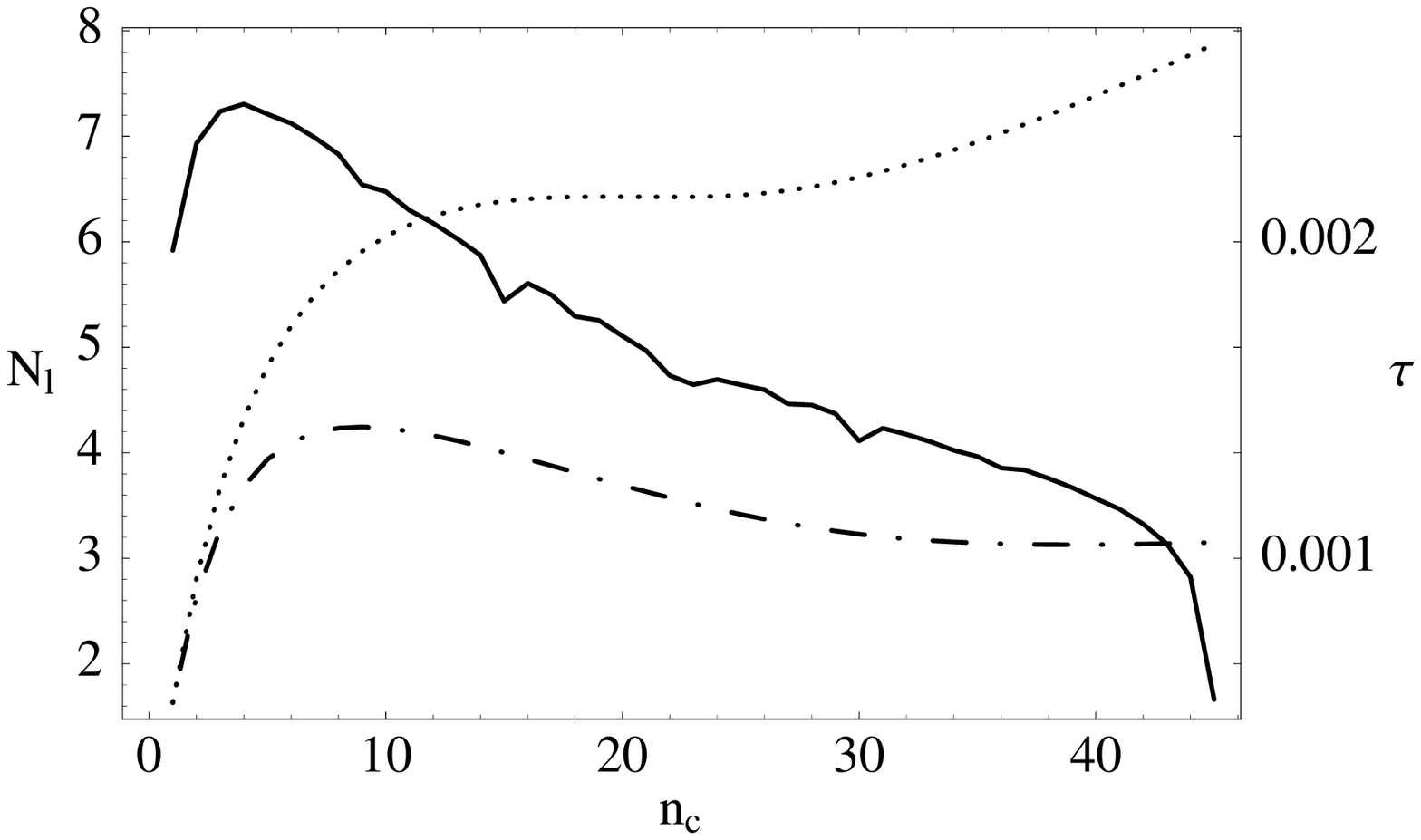}
\includegraphics[width=7.8cm]{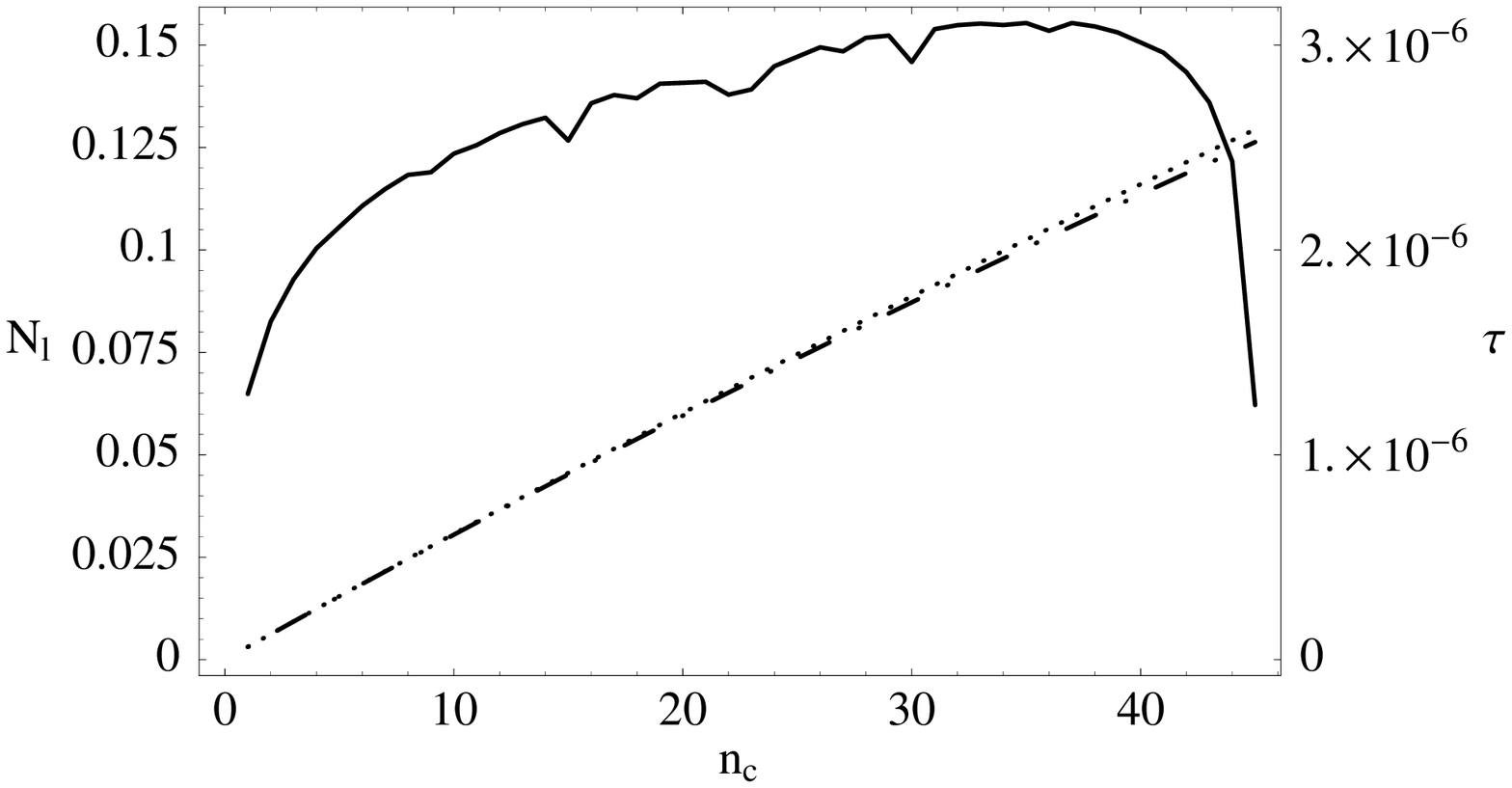}
\caption{For a regular bipartite graph of $n=90$ harmonic oscillators the logarithmic
  negativity $N_l$ is plotted versus the number of connections per
  oscillator $n_c$ (solid line, left y-axis). The dotted and
  dot-dashed line are the plot of $\tau^{1:2\dots n}$ and
  $\sum_{j=2}^{n}\tau_l^{1:j}$ respectively (right y-axis). From top
  to bottom the coupling constant is $\alpha = 1,0.1,0.01$.}
\label{fig:mngInB}
\end{figure}

The results are reported in Fig.~\ref{fig:mngInB} for bosonic
oscillators in a regular bipartite graph (see Sec.~\ref{g:bip}).  We
see that when the residual entanglement begins to be comparable to the
sum of the two-mode tangles, then the entanglement between sets $A$
and $B$ starts to be suppressed.  In other words, when a complex
entanglement structure appears, then particles belonging to the same
set start to become quantum correlated. As a consequence of monogamy,
they can then share no more a high amount of correlations with the
particles belonging to the other set, and the bipartite entanglement
between $A$ and $B$ is suppressed. Notice that as the strength of the
coupling between the oscillators increases the effects of monogamy
become more and more important, as one may expect by general
considerations. The additional frustration in proximity of the fully
connected graph can be instead attributed to symmetric reasons. More
specifically, in a fully connected graph the state of the system is
completely symmetric, which in turn implies a reduction of the
effective size of the Hilbert space associated to the $n$ oscillators.
As a consequence, the entanglement in the system is frustrated too.

\begin{figure}[h!]
\includegraphics[width=7cm]{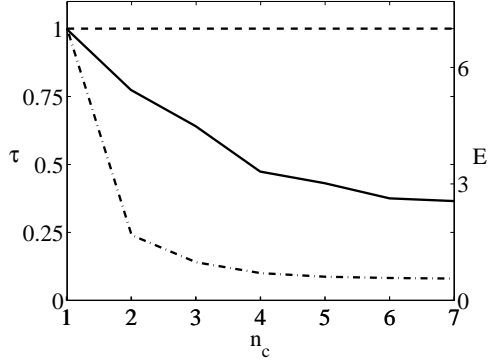}
\caption{For a bipartite graph of XX spin systems with $n=14$,
  the entropy of entanglement $E$ is plotted versus the number of
  connections $n_c$ (solid line, right y-axis). The dashed and
  dot-dashed line are the plot of $\tau^{1:2\dots n}$ and
  $\sum_{j=2}^{n}\tau_l^{1:j}$ respectively (left y-axis). The average
  is taken over 100 realizations.}
\label{fig:mngIn}
\end{figure}

For spin systems we considered the monogamy inequality presented in
Refs.~\cite{CKW00,OV06} where $\tau$ is now the square of the
concurrence, or tangle, a measure of entanglement between qubits.
Fig.~\ref{fig:mngIn} shows, for the regular bipartite graph with $n_c$
connected neighbors, the quantities $\sum_{j=2}^{n}\tau_l^{1:j}$ and
$\tau^{1:2\dots n}$. For any value of the connectivity, any spin is
maximally entangled with the rest, since $\tau^{1:2\dots n}=1$.
However, as the number of connections increases the structure of
entanglement becomes highly non-trivial. This, as said, limits the
bipartite entanglement between sets $A$ and $B$. Notice that, due to the
absence of an on-site interaction in \refeq{HXY}, an analogy can
be seen between the XX system and the harmonic one in the 
strong-coupling regime.

\section{Conclusions}\label{esco}
We have analyzed the interplay between ground-state entanglement and
the connectivity in spin-$1/2$ and bosonic systems with two-body
interactions. We have shown that the ground-state entanglement does
not necessarily increase by introducing more interacting terms in the
Hamiltonian. Actually, for some systems, it does decrease with the
number of connections. From a more applied point of view, the amount
of entanglement across different bipartitions of a system has been
related to the complexity of the classical description of the state
\cite{V03}.  Although here we just focus on symmetric partitions, it
seems reasonable to expect that such partitions have the highest
entanglement (among the contiguous ones). Thus, the previous analysis
suggests that the bipartite entanglement for many highly connected
systems is similar to the one for low-connected Hamiltonian systems,
where efficient classical algorithms exist, like for example Density
Matrix Renormalization Group (DMRG). As a matter of fact, the latter
method has recently been efficiently applied to a specific highly
connected model, in order to analyze quantum phase transitions in spin
glasses \cite{RS06}. The efficiency of DMRG in this scenario is not
trivial and it is related to the fact that the specific system
analyzed in Ref.~\cite{RS06} turns out to be only slightly entangled.
We have shown here that a large variety of systems may have such a
character, due to the fundamental constraint imposed by the monogamy of
entanglement. Our results, then, may encourage the search for novel
classical algorithms able at simulating highly connected quantum
systems.

\acknowledgments We acknowledge S. Iblisdir and R. Or\'us for
discussion. This work is supported by the EU QAP project, the Spanish
MEC, under the FIS 2004-05639 project and the Consolider-Ingenio 2010
QOIT program, the ``Ram\'on y Cajal" and ``Juan de la Cierva" grants,
the Generalitat de Catalunya, and the Universit\'a di Milano under
grant ``Borse di perfezionamento all'estero".




\begin{thebibliography}{999}
\bibitem{note} This is due to the fact that in both cases the matrix
  $V''F$ is semidefinite (see Ref.~\cite{AEP+02} for details). For the
  sake of completeness, we recall that $V''$ is defined by considering
  the bipartition $V=\bigl( \begin{smallmatrix} V'&V''\\ V''&V'
\end{smallmatrix} \bigr)$, whereas $F_{i,j}=\delta_{i,n+1-j}$.
\end{thebibliography}
\end{document}